\documentclass[preprint,tightenlines,eqsecnum,aps,epsfig]{revtex4}
\usepackage{epsfig}

\def\beq{\begin{equation}}   \def\eeq{
\end{equation}}

\begin{document}
\title{
The increase with energy of the parton transverse momenta in the
current fragmentation region and related pQCD phenomena in DIS at
very high energies.
 \footnote{Talk given by B.Blok at the small x
physics 2008 conference at Columpari, Crete, July 2008.}}
\author{ B. Blok\email{Email :blok@physics.technion.ac.il} }
\affiliation{Department of Physics, Technion---Israel Institute of
Technology, 32000 Haifa, Israel}
\author{ L. Frankfurt\email{E-mail: frankfur@lev.tau.ac.il} }
\affiliation{School of Physics and Astronomy, Raymond and Beverly
Sackler Faculty of Exact Sciences, Tel Aviv University, 69978 Tel
Aviv, Israel}
\author{ M.Strikman\email{\E-mail: strikman@phys.psu.edu}}
\affiliation{Physics Department, Penn State University, University Park, PA, USA}
\thispagestyle{empty}
\begin{abstract}
We find, within the pQCD dipole model for DIS processes, a rapid
 increase with energy of the scale of parton momenta  in the current fragmentation region in the limit $s\to \infty$ and  fixed $Q^2$ .
We explain the equivalence between the dispersion representation over
$Q^2$ for  LT pQCD zero angle amplitude of
$\gamma^*+T\to \gamma^*+T$  scattering at large energies and
the $k_t$ factorization,
 and
use it to evaluate the scales of the hard
processes in the current fragmentation region.   We derive within
the black disc (BD) regime the modified Gribov formula for the
total cross-section of the DIS. We  evaluate the coherence length of the
processes relevant for the BD regime and
find that it increases with energy  as $\sim s^{0.6}$ i.e.
significantly slower than in the parton model ($1/2m_Nx$ - the
Ioffe length) as well as in pQCD. In the BD regime we estimate the
gluon densities,  local in the coordinate space, and find that
they do not decrease with the energy. We discuss
briefly how the new pQCD phenomena may reveal itself in the
proton-proton
collisions at  LHC.
 \end{abstract}

\maketitle \setcounter{page}{1}
\section{Introduction}
\par

\par
The aim of this talk is to
demonstrate  the existence of
the new hard QCD phenomena in DIS at small $x=Q^2/s$
and large but fixed $Q^2$, in the vicinity of the onset of the BD regime, in the kinematic domain where the pQCD is still valid,
 and to apply them to the
physics to be observed at the LHC.   We use the
dipole model and
$k_t$ factorization  \cite{CH,CE} as a tool and find that
in the limit $x_B\to 0$ and large but fixed $Q^2$ the
essential transverse momenta of partons within the dipole are rapidly
 increasing with energy in contrast with  the limit $Q^2\to \infty$
and fixed $x_B$. In
the case of the
longitudinal photon cross-section $\sigma_L$ the essential region
of integration over the transverse momenta of the constituents
within the quark dipole is $k_t^2\approx Q^2/4$ at moderate and
large $x$ but increases with the energy at sufficiently small $x$.
Similar effects occur  in the case of the transverse photon
cross-section $\sigma_T$, where the relevant invariant masses
increase with the energy even more rapidly. This result is in a
sharp contrast with the regime of moderate energies like the
energies of HERA ($s\le 10^5$ GeV$^2$), where dominant configurations in the
cross section $q\bar q$  have invariant masses
$M^2(q\bar q)\le Q^2$.

\par The rapid increase of the characteristic transverse scales in the fragmentation region has been discussed in ref.
\cite{FSW}, however  within
the black disk (BD)  regime.  Our new result is the prediction of the
  increase of  the jet momenta in the fragmentation region, in the kinematical domain where methods of pQCD are still applicable.
To visualize
analytically the origin of this phenomenon
in pQCD we use the double logarithmic approximation which is a rather rough
approximation. The calculations show that in $\sigma_L$ the  relevant
invariant masses of the $\bar q-q$ pairs (dipoles) increase to
$M^2\sim 1.5-2 Q^2$   for the quark dipole-nucleon interactions
at invariant energies  $s=10^5-10^6 GeV^2$.
 Let us stress that this new regime is characterized by the
increase of the parton transverse momenta with energy in the photon
 fragmentation region and thus it is different from the well known
phenomena of the  diffusion to the large transverse momenta in
$\ln k_t^2$ plane in the center of rapidity \cite{BFKL}.

\par We have carried our calculations in the leading logarithmic
approximation. However, we expect that qualitatively our results will
be valid in the NLO and beyond, because to a  large extent the NLO corrections are taken into account in the phenomenological fits that
fix the parameters of the parton distributions in the LO approximation.
Remember that the parton distributions have been extracted from data
in the limit of $Q^2\to \infty$ and fixed x where effects discussed in the paper should be a correction.

\par For the deep inelastic lepton scattering off a nucleon we
evaluate the dependence
on the energy $s$ and transverse momenta $Q^2$
of the three characteristic scales. The first scale is the maximum
in the distribution of jets  in the current fragmentation region of
the total cross-section over $M^2$
-$M^2_1$ . Then we evaluate the region of $M^2$ that
gives
the median and the dominant contributions  in the cross-section.
We first determine the scale $M^2_{t1}$ that
gives $50\%$  of the cross-section, and then determine the characteristic transverse scale that characterizes
the "tail",
 $80\%$ of the total cross-section
 ,it's upper
boundary will be denoted as $M^2_{t2}$. The value of $80\%$ is
chosen
to quantify
$M^2$  that is relevant for the
process. The more detailed analysis  will be given elsewhere.
\par
Another important scale is the scale of the invariant mass of the
dipole pair  $M^2_b$ when the partial width for dipole-nucleon
scattering reaches one for the central impact parameter $b=0$.
This scale characterizes the onset of the black disc limit, and is
close to that
obtained
in the similar analysis in refs. \cite{Rogers1,Rogers2,Mark1}.

\par
As an application of the above discussed results we
find
that at sufficiently high energies  the variety of  coherence lengths
substitutes the Ioffe length -$1/2m_Nx$ \cite{GIP,Ioffe} familiar
from QED \cite{LP,TM}.
 The derived  pattern is different also from the one expected in
the
application of the DGLAP formalism for the scattering
off the photon, where the logs come from the region of
integration over the transverse momenta $0 \ll k^2_t\ll Q^2$.
We pay special attention to  the coherence length that corresponds
to the configurations responsible for the onset of the black disk limit (and dominant in the  total  cross-section).
We show that taking into account  the contribution of dipoles with large
masses and of black disc limit to explain the
probability conservation leads to the coherence length $l_c$
increasing more slowly with the  energy. This is the because the coherence
length for a given process is
\beq
 l_c=s/(M^2(s)+Q^2),
\label{ioffe1}
\eeq
where $M^2(s)$ is the typical M important in the inelastic(total)
cross section. Since effective $M^2$ is  increasing with the energy -see above discussion - the coherence length for the onset of the
black limit is $l_c\le (s/(M^2_b(s)+Q^2))$,  and calculations show that
 this coherence length increases with the energy much slower than Ioffe length, namely as  $l_c\sim s^{0.6}$.
\par
The increase with energy of  the $M^2_b$ scale has been discussed in
detail in \cite{FGMS,FSW,Rogers1,Rogers2}. However, the slowing of the increase of the coherence lengths  with energy in the
kinematics
corresponding to the onset of the BD regime, the  modified Gribov formulae
for the structure functions  in the BD regime are our new results.

\section{The dipole approximation  approach.}
\par
We use in our calculations the dipole approximation, combined with
the $k_t$ factorization. It follows from  the QCD factorization
theorem \cite{FS,FRS} that in the LO approximation inelastic cross
section of  the scattering of the longitudinally polarized virtual
photon off a hadron target is calculable in terms of the
light-cone wave functions of the virtual photon:
\begin{eqnarray}
 \sigma &\sim &\int \frac{d^2 k_t}{2(2\pi)^3}\int d^2r_t dz
 \frac{1}{2(2\pi )^3} \nonumber\\[10pt]
&\times&  \psi(r_t,z)(2\psi (r_t,z)-\psi (r_t+k_t)-\psi
(r_t-k_t))\nonumber\\[10pt] &\times&
\frac{4ImT^{ab}_{\mu_1\mu_2}q^{\mu_1}q^{\mu_2}}{s}.\nonumber\\[10pt]
\label{8.1}
\end{eqnarray}
Here $r_t$ is the transverse  momentum of the constituent within the
dipole and  $k_t$  is the transverse momentum of the exchanged
gluon. The tensor $T^{ab}$ is the sum of the diagrams describing
imaginary part of the amplitude for  the gluon scattering off the target
T. The integral of $T^{ab}$  over $d^{2} k_t$ is proportional to the  gluon
structure function of the target T. The function $\psi$ is the
wave functions of the virtual photon , and $q^\mu$ is  the photon
momentum.
\par
Within the LO accuracy which we use in the analysis this formula
can be rewritten as \beq \sigma\sim \int^1_0dz\int d^2r_t
(\nabla_i\psi(\vec r_t,z))^2x'G(x',4r_t^2).\label{mas}\eeq Here
the derivatives are over $r_{it}$, $M^2=r^2_t/(z(1-z))$ is the
invariant mass of the dipole, $x'=(M^2+Q^2)/s$. The function
$G(x',4r^2_t)$ is the integrated gluon function.  The  above equation
is the generalization of LO DGLAP, and BFKL approximations
to the interaction of the high energety dipole with the wave function given by pQCD,
which accounts for
the $k_t$ factorization theorem. (Within  the DGLAP
approximation the structure function $xG$  in the above formulae
depends on $Q^2,x=Q^2/s$.)

The above equation has rather general justification at small x,
where the LT approximation of pQCD is applicable.  Actually it has been
understood in preQCD period that in a quantum field theory,  where the
coherence length significantly exceeds the radius of  the target T at
large energies, the electroproduction amplitude in the target rest
frame is given by a dispersion integral over $Q^2$ \cite{Gribov}.
The pQCD guarantees another general property: in the target rest frame
description: the smaller size of the configuration in the wave
function of projectile photon leads to a smaller interaction with
the target. This property is ensured by the $k_t$ factorization
theorem for the interaction of sufficiently energetic dipole with
a target.   Both properties together give the LO
eq.\ref{mas} . In
the NLO approximation
the structure of formulae is the same except the appearance of the
additional $q\bar q g,  ...$
 components in the wave function of
photon due to the necessity to account for the QCD evolution of the
photon wave function \cite{FS}.
\par For the cross section of the interaction of the longitudinal photon,
  explicitly differentiating the photon wave function, we find:
 \beq
\sigma_L=\frac{\pi\alpha_{\rm e.m.}\sum e^2_qF^2Q^2\alpha_s}{12}
 \int dM^2
\frac{M^2}{(M^2+Q^2)^4}\cdot x'G(x',M^2).\label{r10}\eeq
\par
Here $F^2=4/3$ for the colorless dipoles build of color triplet
constituents. The masses of the constituents of the dipole were
neglected since we restrict our consideration to the spatially
small dipoles only.
\par
Similarly, for  the pQCD contribution to the cross-section of the transverse
photon we find:
 \beq \sigma_T= \frac{\pi\alpha_{\rm e.m.}\sum
e^2_q F^2\alpha_s(r^2_t)}{12}\int dM^2
\frac{(M^4+Q^4)}{(M^2+Q^2)^4}\cdot x'G(x',4r^2_t).\label{r11}\eeq
Here in the practical use we introduce a cut off $M^2z(1-z)>u_t^2$,
where $u_t$ is a lower cut off, beyond which we can not use the
perturbation theory, and in order to obtain the real full
cross-section we must add the contribution of the aligned jet
model (AJM).
\par
We shall use this equation  for the illustrative calculations only, and check it's usefulness
by
studying
the cut-off dependence. The key difference from the longitudinal
photon case is that the asymmetry in the z configurations
(the aligned jet model)
gives significant contribution into the
transverse cross-section even at relatively high energies
( this contribution dominates in diffractive processes
at HERA energies). These contributions
correspond to large invariant masses even when
the actual transverse momenta are small, since
$M^2=r^2_t/(z(1-z))$, and
large for $z\rightarrow 0,1$. We
expect however that for sufficiently large energies the
relative
contribution of the alligned jet model decreases
since
the contribution of the symmetric pQCD configurations increases
with energy much
more rapidly. Indeed, the contribution of the low $r_t$ is
multiplied by a structure function at the second argument of order
$4r^2_t$.

\par  The formulae  discussed above in the momentum space obtain
the transparent form  in  the coordinate space description. In the
leading order pQCD, the cross-section of the interaction of the
longitudinal and transverse photons off the nucleon have the form
:
 \beq
 \sigma_L(T)(x,Q^2)=\int^{1}_0dz\int d^2d
\sigma(d,s,Q^2)\vert \psi_{\gamma L(T)}(z,d)\vert^2. \label{1.1} \eeq
Here and below $d$ is the transverse diameter of the dipole, and
$\psi$ is the light-cone wave function of the photon and  $\sigma$
 is the cross section for the quark dipole scattering off target T.
\par

The connection between $d^2$  and transverse momenta and invariant
masses is given by $d^2=\pi^2/(4r^2_t)$, $ M^2=r^2_t/z(1-z)$
\cite{FKS}.
 The cross-section of the dipole -hadron
interaction is given by the equation \cite{BBFS,FMS} : \beq
\sigma(d,x')=F^2(\pi^2/3)d^2\alpha_s(M^2)x'G(x',M^2). \label{1.3}
\eeq Here $d$ is the transverse scale of the dipole, and once
again $x'=(M^2+Q^2)/s$. The derivation of this formula heavily uses validity of the LO $ k_t$
  factorization theorem -see the above discussion.

 The wave functions of the longitudinal and
transverse dipole are given by \beq \vert
\psi_L(z,d)\vert^2=(6/\pi^2)\alpha_{e.m.}\sum^{n_f}_1e^2_qQ^2z^2(1-z)^2K_0^2(\epsilon
b),\label{1.2}\eeq where $\epsilon^2=Q^2z(1-z)$ ,and
 \beq
\vert\psi_T(z,d)\vert^2=(3/2\pi^2)\alpha_{e.m.}\sum^{n_f}_1e^2_q
(z^2+(1-z)^2)(\epsilon^2K^2_1(\epsilon d).\label{2.4}\eeq Here
$K_0,K_1$ are the standard zero and first order
 MacDonald functions.
\par Note that with the LO accuracy the cross-section
representations as integral over transverse momenta and over
dipole sizes are just connected by a direct and inverse Fourie
transformation.
\par In the practical application of the dipole model for the longitudinal photon the above formulae
can be further simplified since it may be shown that the dominant
contribution in the pQCD comes from z=1/2. In this case it is
possible to further improve the model, introducing a parameter
$\lambda$, such that $d^2=\lambda/M^2$, and determining it from
the best fits with the data on HERA structure functions and
$J/\psi$ production. For the symmetric configurations, taking into
account the above connection between $d^2$ and $r^2_t$, we have
$d^2=\pi^2/M^2$. However it is possible to show that the results
do not change for the whole region of $\lambda >4$ \cite{Rogers1},
and thus are insensitive to the precise value of the coefficient
in the formula connecting $d^2$ and $M^2$.
 At high energies a new QCD phenomena appear- the increase with energy of the transverse momenta of constituents
 within the dipole.
This
property is absent in the LO, NLO prescriptions for the
calculation of Feynman diagrams of pQCD  where the contribution
of the kinematical region: $M^2\gg Q^2$ is neglected. In other
words, the LO/NLO are internally inconsistent
at sufficiently large energies since they ignore the basic property
 of pQCD-smaller size of dipole-faster increase of a cross section with energy.
We explained above that this
generalization is necessary to account for basic property of pQCD
that the interaction of color neutral dipole of zero transverse
size with a target should be 0.

\par The next practical question is how to parameterize the structure
functions for the realistic calculations . Here we shall use
CTEQ5L  distributions \cite{STEQ}. The CTEQ5L distributions has
been shown to be in good agreement with HERA data \cite{Rogers2},
in the range  of $x\sim 10^{-3},10^{-4}$. To extrapolate to very
small x, we shall use the approximate formulae in the form:
 \beq
xG\sim a(M^2)/x^{c(M^2)}. \label{n1} \eeq Here the functions \beq
a(M^2)= 2.00123 - 1.69772\cdot 10/M^2 + 3.07651/\sqrt{M^2/10.} -
0.228087\cdot \log{M^2/10.}, \label{n2} \eeq \beq c(M^2)=0.045\log
(M^2)+0.17, \label{n3} \eeq where $M^2$ is in GeV$^2$.
 This formula is the fit to the observed
 behavior of the structure functions in HERA for $150 {\rm
GeV}^2\ge Q^2\ge 3$ GeV$^2$ , made by ZEUS and H1 collaborations
\cite{H1}.
  We shall use this function also in the kinematics where
the CTEQ parton distributions are  absent. Note that observed
increase of $c (M^2)$ with $M^2$ can be explained as due to $Q^2$
evolution of parton distributions cf. ref.\cite{DDT} and within
the resumation approach \cite{Ciafaloni,ABF}.
\section{The  hard fragmentation processes.}

\par
 We have carried both analytical and numerical analysis of the
 transverse scales. The analytical calculation
has been made for the toy model based on the double logarithmic
approximation  and will be published in the full version of the
paper. The numerical results will be presented below. We shall
extrapolate our results to energies of order
$s\sim 10^7 $ GeV$^2$.
These energies are unattainable due to the fact that there is no
e-p collider for such energies. (The proposed e-p DIS facility at the LHC may reach the energies of order $ 10^5$ GeV$^2$. However
these results are interesting both from the theoretical point of
view (probing the limits of the pQCD at very high energy), and
from the practical point of view,  giving the information
 about the parton distributions at the LHC. The relation of our
 results to the processes at the LHC will be
discussed in section V.
\subsection{Numerical analysis: longitudinal photons.}
\par
Let us  analyze the integrand in the integral  representation of
the cross-section \ref{1.1} as the integral over  $d^2$. This
integral can be rewritten as integral over $M^2=4/r^2_t$,
$r^2_t=(\pi^2/4d^2)$ \cite{FKS}, and we use the dominance of the
symmetric configurations $z=1/2$. This approximation is good since
$z(1-z)$ is a slow function of z for a large range of z. With the
logarithmic accuracy such integral must numerically be the same as
the integral over $M^2$ in the momentum representation of the
corresponding cross-section. We carried our calculations both in
the coordinate and momentum representations and found that the
results coincide within a  $10\%$ accuracy.
\par
Let us first consider the quantity $d\sigma_L/d M^2$. This
quantity in the z=1/2 approximation is proportional to the
transverse momentum $r_t$ jet distribution in the fragmentation
region.

 In order
to characterize the  behavior of the density we calculate two
scales: $M^2_1$ and $M^2_t$. The first of these scales
characterizes the maximum of the density. This scale lies
$M^2_1\le Q^2$, for all the values of the energy less than $10^{11}$
GeV$^2$. In line
 with the expectations based on the toy
model we see that this maximum however slowly increases with
energy for every value of $Q^2$, with the energy dependence for
$Q^2\ge 5 $ GeV$^2$ being approximately $s^{0.04}$.
\par The scale $M_1^2$ however does not give a full
characterization of the density, the reason is that the rapid
decrease of the square of the derivative of the wave function is
compensated partly by the rapid increase of the structure function
with the increasing $M^2$. In order to characterize this effect it
is natural to determine the new scale $M^2_{t1}$ that corresponds
to the cut off in $M^2$ that gives, say, $50\%$ of the total
cross-section, and another scale $M^2_{t2}$ that corresponds to
$80\%$ of the cross-section and characterizes the magnitude of the
tail.
 \par The determination of $M^2_{t1}$ is given in Table 1 for
the characteristic virtualities $Q^2=5,10,20,40,100$ GeV$^2$ and for
realistic energies (up to those achievable at LHC). We see that
for not very small $Q^2$ (starting from $Q^2\sim 10 $ GeV$^2$) and
for HERA energies this scale lies beyond $0.75Q^2$, where $Q^2$ is
an external virtuality, thus justifying the use of DGLAP at these
energies.  We see from the tables that this median scale rapidly
increases with energy, and will overcome $Q^2$ at $Q^2=10-20$
GeV$^2$ already at LHC energies and may significantly overcome
$Q^2$ if we shall go beyond LHC.
\par Let us consider the second scale, $M^2_{t2}$. We see that the
tail exists even at HERA energies for $Q^2<200 $ GeV$^2$, when the
$M^2_{t2}$ exceeds $Q^2$. The tail also increases with the
increase of energy, even more rapidly then the median scale
$M^2_{t1}$. \par The scale  $M^2_{t2}$ is always larger than
$Q^2$. If we extrapolate to the very large (though unrealistic)
energies we see that  it exceeds $Q^2$ for very small $Q^2\sim 5
GeV^2$  by a factor of 5. In other words,
 for sufficiently small momenta
 and high energies the leading logarithmic approximation leads to the very long
  tails, that are strictly  beyond the control of usual
prescriptions for DGLAP and BFKL  approximations and must be taken
into account using $k_t$ factorization theorem, at least in the LO
approximation.
\par  We conclude that for sufficiently small external
virtualities and high energies (large 1/x) one cannot use the
naive DGLAP approximation where transverse momenta in the upper
rung of the ladder are $\ll Q$, while the the dipole approximation
provides a reasonable description of
 the cross-sections. In this approximation the
 LO
 DGLAP/resummed ladder gives the cross-section of the dipole
scattering of the target.
\subsection{The evaluation of the transverse scale: transverse
photons.}
\par
The same two scales $M^2_t$ and $M^2_1$ that we defined for
longitudinal photon can be defined for the transverse photon.
\par In this case however the use of the pQCD expression poses a
problem due to a potential large contribution of the
nonperturbative physics, described by the aligned jet model.
Another difference from the longitudinal photon case is the
large contribution of the asymmetric configurations,
with $z\ne 1/2$. These contributions are dominant in the AJM
model, and
lead to large invariant masses even for small
transverse momenta, due to relation $M^2=r^2_t/(z(1-z)$. In this
talk we are concerned only with the contribution of the pQCD. In
order to exclude the contribution of the AJM (nonperturbative QCD)
we impose
a cutoff  $M^2z(1-z)>\Lambda^2$, which removes the
contribution of the low transverse momenta. The use of the pQCD
formulae, even for illustrative purposes, would be justified only
 if the cross section depends weakly on the cutoff.
We see however, that for  $Q^2<20$ GeV$^2$ even for
the energies of the
order $10^7$ GeV$^2$ there  is a strong dependence
(of the order 20$\%$) of the maximum of the curve $d\sigma/dM^2$
on a very small change of the cutoff  (from $\Lambda=0.75 $
to $\Lambda =1$ GeV$^2$. The data for the median and the tail transverse
scales $M^2_{t1}$, $M^2_{t2}$ are presented
in Tables 3, 4. We see
that for HERA energies there is a dependence of order $10\%$ of
these scales on the cut off even at $Q^2\sim 100$ GeV$^2$. This
dependence however decreases with  energy.
\par The detailed analysis of the relative contribution of the
pQCD and aligned jet model will be given in ref. \cite{BFS}. In
tables 4, 5 we presented our results for transverse scales at
$Q^2=20,40,100$ GeV$^2$. We considered the integrand for
cross-section as a function of the invariant mass and cut off,
after integrating over permitted z for given invariant mass and
cut off.
\par The scale $M_1^2<Q^2$ for all energies and
$Q^2>10$ GeV$^2$ behaves similar to  the longitudinal photon,
However this scale is shifted to smaller invariant masses relative
to longitudinal photons. For every given virtuality the scale
continues to increase with the increase of energy, with
approximately the same rate as for longitudinal photons.
\par The characteristic feature of the invariant mass distribution
for the transverse photons is however that it becomes much
broader, and increases with energy more rapidly than for
longitudinal photons. As a result
although the
transverse scales for HERA energies are slightly smaller than for
longitudinal photons, for high energies, achievable at LHC they
already overcome them. The rate of increase is $M^2\sim s^{0.1}$
for LHC energies and $Q^2\sim 50 $ GeV$^2$.
\par It is once again instructive to look where  $Q^2=M^2_t$ as
a function of energy. We see that the tail starts (with
logarithmic accuracy) for higher $Q^2$ than for the case of the
longitudinal photons.
\par Since our results show qualitative stability with the change
of the cut off and the dependence of cut off (and AJM contribution
) decrease with $Q^2$ and energy, we expect that our results
remain qualitatively the same if we shall take into account the
AJM contribution explicitly.
\par Let us note that the appearance of the tails at sufficiently
large energies (small x) can be expected from the comparison with
the Gribov formulae for the BD regime\cite{Gribov}. Indeed,
assuming a smooth matching of the pQCD and BD regime  the
perturbative distribution must match the black limit spectral
density , and this density leads to
the tail with masses increasing with energy.
Thus the existence of
a large mass tail beyond the naive DGLAP approximation seems to
be  a necessary property of pQCD near the black limit.
\par Finally, we expect that our results will not change in the
NLO/resumed model. The reason is that NLO effects are partly taken
into account in the gluon distribution fits to experimental data.
\subsection{The evaluation of the black scale.}

We can now combine the constraint for the partial wave at zero
impact parameter to become unit with the CTEQ5 structure function
to determine the invariant masses that correspond to the onset of
the black limit.
\par The typical dependence of
 the black limit onset scale $M^2_b$  on energy is presented
in Table 3. for the  gluonic dipole. We consider two cases: one
when the partial wave $\Gamma $ at the central impact parameter
reaches 1, another when it reaches 1/2 \cite{FSW}. Indeed, when
the partial wave reaches 1/2 the probability of inelastic
interactions reaches 3/4, i.e. interactions become strong and pQCD
can not be used any more \cite{FSW}.
\par The characteristic transverse momenta that corresponds to the black scale for the gluonic dipole for the energies
$s=10^6-10^7$ GeV$^2$ are $2-2.5$ GeV/c if we impose
$\Gamma =1$ condition or $4-4.5$ GeV/c if we impose $\Gamma
=1/2$. Note that for HERA energies $(s=10^4$ GeV$^2$ we do not
obtain the black limit for gluonic dipole for $Q^2>40 $ GeV$^2$,
while for lower values of $Q^2$ we obtain formally $k_t\sim 1$
GeV, the value that seems beyond the limits of the applicability
of the method we used here to determine this scale.

\par For fermionic dipoles we see that
 there is no black limit for HERA energies, although nonperturbative effects (corresponding
to $\Gamma=1/2$) seem to start to appear for $k_t<1 $ GeV.  For
the  energies $s\sim 10^6-10^7$ GeV$^2$ the transverse scale is
$\sim 2-2.5$ GeV for the partial wave $\Gamma =1$ condition,  $
3-3.5$ GeV respectively for $\Gamma =1/2$ conditions. These
results are in good agreement with the previous determination of
these scales \cite{Rogers1,Mark1,FSW}.
\par For our purposes it will be important to determine
 the rate of increase of the $M^2_b$. This rate does not depend
on the type of the dipole (fermionic or gluonic) or on the partial
wave condition.
\par
For the gluonic dipole in DIS for s up to $s\sim 10^7$ GeV$^2$ the
black limit scales increase relatively slow as $s^{0.3}$ .
However at the energies above $s\sim 10^7$ GeV$^2$
 the
increase speeds up with $M^2\sim s^{0.4}$ starting from $s\sim
10^8$ GeV$^2$ energies ( for the exponent for partial wave 1/2 we
obtain 0.39, for the exponent with partial wave equal to 1 we
obtain 0.38), and small $Q^2<40 $ GeV$^2$. Note that this region
is where we expect that there is no dip influence.

\par We  show that for moderately large $Q^2$, where the coupling constant is small the cross-section will be dominated by the black
limit.
\par We found from this subsection that for the DIS processes at
very high energies there exist 3 regimes. First the
nonperturbative black disk regime. This regime is valid for
invariant masses up to $M^2_b$, then
the standard pQCD regime described by DGLAP and then the new pQCD regime, where one can not use DGLAP, BFKL  resummed models directly,
but must combine them with the dipole
approximation. In this new regime
a virtual photon
creates
a parton pair with invariant mass larger than the
external virtuality, that scatters off the target by the
DGLAP/resummed ladder.
Thus in
the s$-Q^2$ plane we
find three regions: the black limit region-the
virtualities where the cross-section is dominated by the black
limit, the region where the cross-section is dominated by the
usual pQCD, and the new regime area where the tail gives a
significant contribution to cross-section.
\subsection{A hadron structure function in the black disc limit}

 \par The above formulae make possible first quantitative evaluation
of the numerical  coefficient in the asymptotical expression for the
structure function of DIS  in the black disc limit suggested in
ref. \cite{PRL,AFS}.

The asymptotic behavior of  the hadron structure functions at
$s\to \infty$ in QCD
is rather close to Gribov formulae for the total cross section of DIS derived long ago for heavy nuclear target in the black disc
limit \cite{Gribov}. In the case of the transverse photon
 \beq
 F_2=2\pi R_T^2(s) Q^2 \kappa \rho(M^2\to \infty) (\ln(x_0/x)),
\label{x1}
 \eeq
where  $\rho$ is  the spectral density for  the transition
$e^+e^-\rightarrow {\rm hadrons}$ :
 \beq \rho(M^2)=\frac{\sigma(e^+e^-\rightarrow {\rm hadrons})}{\sigma(e^+e^-\rightarrow \mu^+\mu^-)}.
\label{sdf}
\eeq
 The density $\rho(M^2)$  tends to constant at large invariant masses.
$R_T$ is the maximal impact parameter, where the partial wave reaches 1.  Fourie transform of two gluon form factor of a nucleon:
$F=1/(1+q^2/\mu^2)^2$ measured in the hard diffractive processes
at HERA and at FNAL  describes dependence of a partial amplitude $f$ on the impact parameter
 b : $f(b,x) \propto (1/x)^{\lambda} exp ^{-\mu b}$ .
Here $\mu^2\approx 1 GeV^2$ . In the LT approximation
the structure function of a hadron in DIS is usually parametrized as
$F_2\sim (1/x)^{\lambda}$. Finally   we obtain:
 \beq
  R_T=\lambda(M_b^2) ln(x_0/x)/\mu
 \eeq
  Here $x_0\sim 10^{-2}$,
A slightly more complicated reasoning which uses completeness over the hadronic states shows that Eq.\ref{x1} is valid even if  to
account
for the gluons and quark-antiquarks pairs in the wave function of the dipole.  The  coefficient $\kappa$ in eq. \ref{x1} is determined
 by the upper limit of the integration  over $M^2$ and
 is equal to  $M^2_b\sim s^\kappa$.
Our results show that this coefficient weakly depends on energy,
with $\kappa =0.3-0.4$ at the LHC energies and beyond.

\section{The coherence  length}
\par In the previous sections we determined the absolute values
and the  energy dependencies of the effective transverse scale and
the black limit scale at high energies which allowed us to
evaluate coherence length.The coherence length $l_c$ corresponds
to the time, such that the dipole fluctuation exists this time interval  at a given energy. The
original
suggestion of the existence
of the coherence length in the deep
inelastic scattering was first made by Ioffe, Gribov and
Pomeranchuk
\cite{GIP,Gribov}
It was found already in the sixties by
Ioffe \cite{Ioffe} that the coherence length at moderate $x_B$
is $l\sim 1/2m_Nx_B$ i.e. it linearly increases with energies.
At higher energies we have a whole variety of local coherence
lengths. Here we consider the coherence length, that corresponds
to configurations near the onset of the black limit that dominate
the cross-section at very high energies.
\par Since near the onset of the black limit $M^2\sim s^{0.4}$ we conclude,  that in the vicinity of black limit i.e.
 for sufficiently high energies this coherent length increases like $l_c\sim s^{0.6}$, i.e. it increases with
  energy much
slower than for lower energies, when Ioffe result is applicable.
\par
The important characteristic of the hard processes is the local
gluon density produced in  the hard high energy process. Rough
estimate of
the three dimensional space density of gluons
(number of gluons in a unit of the space volume) gives
 \beq n\sim
N/(l_c \pi /r_t^2) > xG(x,4r^2_t)/(l_c\pi/r^2_t) , \eeq where N is
a total number of the emitted gluons. 
 \par  At  the LHC energies the
coherence length increases with energy as $\sim s^{0.6}$ for sufficiently small $Q^2$ ($\sim
s^{0.55}$ near the energies of order $10^{11}$ GeV$^2$ ). Using
known dependence of the gluon structure function on energy, we
obtain that the number of gluons increases like  $s^{0.4}$,
while the same formulae extrapolated to superhigh energies gives
$s^{0.5}$. This means, that the naive estimate gives slowly
decreasing gluon number density $n\sim s^{-0.2}-s^{-0.1}$.
Here, however,  we neglected the increase of the transverse
momenta that may also contribute to the increase of a local gluon
density.
The realistic estimate of this increase is however impossible in
the LO approximation. Consequently,
 within our accuracy we cannot determine whether the
local (three-dimensional) gluon density is increasing or slowly
decreasing. To address this question more quantitatively we have
to go beyond the naive estimate and perform a detailed calculation
of the local density, taking into the account the inhomogeneity of
the ladder. Such analysis will be done elsewhere.

\par We conclude that there are indications that the local gluon density may
appear large in the black disc limit regime.

The opposite example, of high light cone densities and low local density rapidly decreasing with energy  was considered some time
ago by Mueller
in QED
\cite{MuellerQED}. However in that case
there is no
rapid increase of interaction with energy.

It is worth noting that
discussed above
pattern for the energy dependence of the coherence length
leads to a change of the
structure of the fast hadron wave function as compared to the Gribov picture where the longitudinal size of the hadron
is
determined by the wee parton cloud and
 energy independent
 $\sim 1/\mu$
where $\mu$  is the soft scale. On the other hand a slower rate of the increase of the
 coherent length with energy than $1/m_Nx$ leads to decrease of the
 longitudinal size of the hadron with energy. The typical size is determined by
 the BD momentum at a given impact parameter for a particular energy. Moreover
  since the BD momentum is larger for small impact parameters the nucleon has a form of
 a concave lens. It is of interest also that for the zero impact
   parameter the longitudinal size of a  heavy nucleus is  smaller than for a nucleon.

\section{Experimental consequences}

\par The current calculations of
cross-sections of hard processes
 at the LHC are
based on the use of the DGLAP parton distributions and the
application of the factorization theorem. Our results imply that
the further analysis is needed to define the kinematic regions
where one can use DGLAP distributions. We
showed in the paper
that for DIS
at high energies there are kinematic regions where one is forced
to use a $k_t$ factorization and the dipole model instead of
direct use of DGLAP. A similar analysis must be made for the
pp collisions at
LHC. This
analysis is however more complicated since the proton can not be
approximated by a dipole
and thus the DIS results for the same energies and external virtualities
can not be transferred directly to $pp$ case. This is a problem for
a future work.
\par
The hard processes initiated by the real photon can be directly
observed in the ultrapheripheral
 collisions \cite{Mark1}. The
 processes where
 a real photon scatters on a target,
 and creates
 two
 jets with an invariant mass $M^2$, can be analyzed in the dipole model by formally putting
 $Q^2=0$, while $M^2$ is an invariant mass of the jets. Then with the good accuracy the spectral density
discussed above will give the spectrum of jets in the
fragmentation region. Our results show that the
 the jet distribution over the transverse momenta will be broad
 with the maximum moving towards larger
transverse momenta with increase of the energy and centrality of
the $\gamma A$ collision.

\par Finally, our results can be checked directly, if and when the
LHeC facility will be built at CERN.

\newpage

{

 \centerline{Table 1. The scale $M^2_{t1}$ ($50\%$ of the total cross-section)  for longitudinal photons in
 DIS}

\begin{tabular}{|l|l|l|l|l|l|}\hline
 &$Q^2=5$ GeV$^2$&$=10$ GeV$^2$&$=20$ GeV$^2$&$=40$
 GeV$^2$&$=100$ GeV$^2$\\  \hline
 $s=10^4$ GeV$^2$& 6.5 GeV$^2$&9 GeV$^2$ &16 GeV$^2$&28 GeV$^2$ &60 GeV$^2$\\ \hline
 $s=10^5$ GeV$^2$&7 GeV$^2$&10.5 GeV$^2$&17.5 GeV$^2$&31 GeV$^2$&68 GeV$^2$\\ \hline
 $s=10^6$ GeV$^2$&7.5 GeV$^2$&11 GeV$^2$&19 GeV$^2$&34GeV$^2$& 77GeV$^2$\\ \hline
 $s=10^7$ GeV$^2$&8 GeV$^2$&12 GeV$^2$&21 GeV$^2$&37 GeV$^2$&87 GeV$^2$\\  \hline
 \end{tabular}

\centerline{Table 2. The scale $M^2_{t2}$ ($80\%$ of the
cross-section )for longitudinal photons in
 DIS}

\begin{tabular}{|l|l|l|l|l|l|}\hline
 &$Q^2=5$ GeV$^2$&$=10$ GeV$^2$&$=20$ GeV$^2$&$=40$
 GeV$^2$&$=100$ GeV$^2$\\  \hline
 $s=10^4$ GeV$^2$&10.5 GeV$^2$&17 GeV$^2$ &30 GeV$^2$&54 GeV$^2$ &120 GeV$^2$\\ \hline
 $s=10^5$ GeV$^2$&11.5 GeV$^2$&19 GeV$^2$&34 GeV$^2$&60 GeV$^2$&140 GeV$^2$\\ \hline
 $s=10^6$ GeV$^2$&12.5 GeV$^2$&21 GeV$^2$&38 GeV$^2$&67GeV$^2$&160GeV$^2$\\ \hline
 $s=10^7$ GeV$^2$&14 GeV$^2$&23 GeV$^2$&42 GeV$^2$&75 GeV$^2$&180 GeV$^2$\\  \hline
 \end{tabular}

\centerline{Table 3. The scale $M^2_{t1}$ for transverse photons
in DIS.}

 \begin{tabular}{|l|l|l|l|}\hline
 &$Q^2=20$ GeV$^2$&$=40$
 GeV$^2$&$=100$ GeV$^2$\\  \hline
 $s=10^4$ GeV$^2$&24(20) GeV$^2$&31(26) GeV$^2$ &50(45) GeV$^2$\\ \hline
 $s=10^5$ GeV$^2$&27(23) GeV$^2$&37(33) GeV$^2$&59 (55)GeV$^2$\\ \hline
 $s=10^6$ GeV$^2$&32(28) GeV$^2$&43(38)GeV$^2$&73 (70)GeV$^2$\\ \hline
 $s=10^7$ GeV$^2$&38(33) GeV$^2$&52 (48) GeV$^2$&90 (86) GeV$^2$\\  \hline
 \end{tabular}

 \centerline{Table 4. The scale $M^2_{t2}$ for transverse photons in
 DIS}

 \begin{tabular}{|l|l|l|l|}\hline
 &$Q^2=20$ GeV$^2$&$=40$
 GeV$^2$&$=100$ GeV$^2$\\  \hline
 $s=10^4$ GeV$^2$&73 (65) GeV$^2$&100 (90) GeV$^2$&160 (150) GeV$^2$\\ \hline
 $s=10^5$ GeV$^2$&83 (75) GeV$^2$&120 (110) GeV$^2$&200 (190) GeV$^2$\\ \hline
 $s=10^6$ GeV$^2$&96 (88) GeV$^2$&140 (130) GeV$^2$&260 (250) GeV$^2$\\ \hline
 $s=10^7$ GeV$^2$&110 (100) GeV$^2$&180 (170) GeV$^2$&330 (320) GeV$^2$\\  \hline
 \end{tabular}

\centerline{Table 5. The scale $k_t$ for the onset of the black
disk
 regime.}

 \begin{tabular}{|l|l|l|}\hline
 &$\Gamma =1$&$\Gamma=1/2$\\  \hline
 $s=10^4$ GeV$^2$& 1 GeV&1.6 GeV\\ \hline
 $s=10^5$ GeV$^2$&1.6 GeV&2.3 GeV\\ \hline
 $s=10^6$ GeV$^2$&2.3 GeV&3.2 GeV\\ \hline
 $s=10^7$ GeV$^2$&3.3 GeV&4.5 GeV\\  \hline
 \end{tabular}

\bigskip
\bigskip
The values of $M^2$ here in the tables 3,4 correspond to $r_t$ cut
offs 1 GeV, and 0.75 GeV (in the brackets)
\end{document}